\documentclass[letterpaper, 10 pt, conference]{ieeeconf}  

\IEEEoverridecommandlockouts                              
\usepackage{amsfonts}
\usepackage{eucal}
\usepackage{amsmath}

 \usepackage{amsthm}
\usepackage{cite}
\usepackage{graphicx}
\usepackage{multirow}
\usepackage{array}
\usepackage{url}

\title{Technical Report: Distribution Temporal Logic: Combining Correctness with
Quality of Estimation}
\author{Austin Jones, Mac Schwager, Calin Belta}
\date{\vspace{-11pt}}
\usepackage{algorithm}
\usepackage[noend]{algorithmic}

\newcommand{\LTLUNTIL}{\ensuremath{\mathcal{U}}}
\newcommand{\LTLNEXT}{\ensuremath{\bigcirc}}
\newcommand{\LTLEVENTUALLY}{\ensuremath{\ \diamondsuit\ }}

\newcommand{\ARUN}{\ensuremath{r}}

\newcommand{\FUNCTION}[2]{\STATE \textbf{function} #1(#2)}

\theoremstyle{plain}

\theoremstyle{definition}
\newtheorem{dfn}{Definition}
\newtheorem{prob}{Problem}

\newtheorem{example}{Example}
\graphicspath{{../../../Images/}}

\begin{document} 

\maketitle 
\let\thefootnote\relax\footnotetext{Austin Jones is with the Division
  of Systems Engineering, Mac Schwager and Calin Belta are with the
  Division of Systems Engineering and the Department of Mechanical
  Engineering at Boston University, Boston, MA 02115. Email:
  \{austinmj,schwager,cbelta\}@bu.edu

This work was partially supported by ONR under grant ONR MURI N00014-09-1051 and by NSF under grant CNS-1035588}

\begin{abstract}
  We present a new temporal logic called Distribution Temporal Logic
  (DTL) defined over predicates of belief states and hidden states of
  partially observable systems.  DTL can express properties involving
  uncertainty and likelihood that cannot be described by existing
  logics.  A co-safe formulation of DTL is defined and algorithmic
  procedures are given for monitoring executions of a partially
  observable Markov decision process with respect to such formulae.  A
  simulation case study of a rescue robotics application outlines our
  approach.
\end{abstract}

\section{Introduction}
\label{intro}

Temporal logics (TLs) provide a rigorous framework for describing
complex, temporally ordered tasks for dynamical systems.  Temporal
logic formulae can be used to describe relevant properties such as
safety (``Always avoid collisions''), reliability (``Recharge
infinitely often''), or achievement (``eventually reach
destination'')\cite{Baier2008}.  In this work, we define Distribution
Temporal Logic (DTL), a new kind of TL for specifying tasks for
stochastic systems with partial state information.  The logic is
well-suited to problems in which state uncertainty is
significant and unavoidable, and the state is estimated on-line.  Many such systems arise in
robotics applications, where a robot may be uncertain of, for example,
its own position in its environment, the location of objects in its
environment, or the classification of objects (e.g. `target' or
`obstacle').

We represent the system as a Partially Observable Markov Decision
Process (POMDP), and we update a Bayesian filter on-line to give a
current probability distribution over the hidden state.  This
probability distribution is itself treated as a state, called a
\emph{belief state} in the POMDP literature.  We define DTL
over properties of belief states as well as hidden states.  
With DTL, we can describe
such tasks as ``Measure the system state until estimate variance is 
less than $v$ while minimizing the probability of
entering a failure mode'' or ``If the most likely card to
be drawn next is an Ace, increase your bet''.  DTL is a promising
framework for high-level tasks over POMDPs as it can be used to
describe the value of taking observations, as well as describing
complex tasks defined over the hidden states of the system, and how to
react to gains in certainty about the state of the
system.


Current research on temporal logic specifications for dynamical
systems can be broadly divided into three common problems of
increasing difficulty: (i) \em monitoring \em whether a single execution of a
system satisfies a TL formula, (ii) \em model checking \em whether some or all
executions of a system satisfy a formula, and (iii) \em synthesis \em of
control policies to ensure formula satisfaction.
Solutions for all three of these problems have been heavily studied
for deterministic systems with various kinds of dynamics.  For
stochastic systems with a fully observable state, some results exist
for all three of these problems as well.  For example, probabilistic
computational tree logic (PCTL) can be used to describe temporal logic
properties of Markov chains \cite{Baier2008}.  The
probability of  satisfaction of a TL formula over Markov chains can be calculated
exactly using a reachability calculation \cite{Baier2008} or estimated from a
subset of sample paths using statistical model checking
\cite{Sen2005}.  Formal synthesis for probabilistic robots modeled as
Markov decision processes is also an active area of research
\cite{Lahijanian2011,Ding2011,Wolff2012}.  In contrast to these works,
our focus in this paper is on stochastic systems with a hidden state.
This paper introduces DTL as a means to formally pose these standard problems
over such systems, and provides monitoring results by giving a procedure to verify
\emph{ex post facto} with what probability a particular execution of a
POMDP satisfies a particular DTL specification.  The more difficult problems
of DTL model checking and synthesis will be investigated in future research.

Formal methods for stochastic systems with hidden states are difficult to
develop.  In general, if TL formulae are defined with respect to hidden states,
their logical satisfaction can only be verified up to a probability based on 
partial knowledge of the state.
The logic POCTL* is an extension of PCTL that
describes TL properties over hidden states and
observations in Hidden Markov Models (HMMs) \cite{Zhang2004,Zhang2005}.
 HMMs are partially observable systems
in which the true (hidden) state of the system evolves according to a
Markov chain and can be probed by a related observation process.  
POCTL* is used for checking properties such as ``The probability that
the sequence of hidden states $s^0 \ldots s^t$ produces an observation
sequence $o^0 \ldots o^t$ is less than 0.1.''  


POMDPs
\cite{Kaelbling1998} are extensions of HMMs in which actions can be
taken to affect the probabilistic evolution of the hidden states and
the observation process.  Recent development of point-based
approximation methods \cite{Smith2005,Shani2012,Pineau2003,Kurniawati2008} and
bisimulation-based reduction methods \cite{Castro2009,Jansen2012} have
made it possible to find sub-optimal solutions for maximizing the
expected reward defined over hidden states in high-dimensional POMDPs
with low computational overhead. It is well known, however, that
maximizing the actual reward gathered in an execution of a POMDP is
undecidable \cite{Madani2003}.  
Synthesizing policies over POMDPs to maximize the probability
of satisfying a TL formula over hidden states is thus a hard problem,
though some results exist for synthesis over short time
horizons \cite{Wongpiromsarn2012} and in
systems where TL satisfaction can be guaranteed \cite{Cimatti1999}.

The best action to take in a POMDP to increase the probability of satisfaction
depends intimately on the quality of knowledge of the system.  POCTL*
describes the quality of knowledge based on the observation process, but
this approach ignores the richness of information conveyed by belief
states.  Information-theoretic measures defined over belief states can
quantify the certainty (i.e. Shannon entropy) of the current estimate
or the expected informativeness (i.e. mutual information) of future
actions \cite{Shannon1948,Cover2006}. Considering these two
measures in mobile robots have increased environmental estimation quality \cite{Julian2012,Bourgault2002,Hoffmann2010,Choi2010}; incorporating 
them into TL-based planning for POMDPs will possibly yield similar results.
Belief states can also be used to select the most likely hypothesis for the current hidden state.  

Our intention in this work is to introduce a new logic to leverage the
richness of information conveyed in the belief state.  Specifically,
our contributions in this work are:
\begin{itemize}
\item We define syntactically co-safe linear DTL (scLDTL), a DTL that
  can be used to prescribe finite-time temporal logic behaviors of
  POMDPs.
\item We demonstrate that DTL can describe behaviors in partially
  observable systems that are not describable by current TLs
\item We provide an algorithmic procedure for evaluating the
  probability of satisfaction of an scLDTL formula with respect to an
  execution of a POMDP. 
\end{itemize}
We intend to extend these results to synthesizing decision policies to
maximize the probability of satisfying a scLDTL formula in future
work.


\section{Preliminaries}
\label{prelims}
For sets $A$ and $B$, $2^A$ denotes the power set of $A$, $A \times B$ is the Cartesian product of $A$ and $B$, and $A^n = A \times A \times \ldots A$.  We frequently use the shorthand notation $x^{1:t}$ for a time-indexed sequence $x^1 \ldots x^t$. The set of all finite and set of all infinite words over alphabet $\Sigma$ are denoted by $\Sigma^{*}$ and $\Sigma^{\infty}$, respectively.

A \em partially observable Markov decision process \em (POMDP)  \cite{Monahan1982,Kaelbling1998,Varakantham2006} is a tuple $POMDP = (S,\hat{p}^0,P,Act,Obs,h)$ where $S$ is a set of (hidden) states of the system, $Act$ is a collection of actions, and $P:S \times Act \times  S\to \mathbb{R}$ is a probabilistic transition relation such that if $POMDP$ is in a state $s$, taking the action $a$ will drive the system to state $s'$ with probability $P(s,a,s')$. After the hidden state evolves, the system generates an observation from the set $Obs$ with probability  $h(s,a,o) = Pr[o $ seen $|  a$ taken, $POMDP$ in state $s]$.  The system maintains a belief state $\hat{p}^t$ of the 
current state of $POMDP$, where $\hat{p}^t(s) = Pr[POMDP$ in state $s$ at time $t| a^{0:t-1}$ taken, $o^{1:t}$ seen$]$, via sequential application of the recursive Bayes filter
\begin{equation} \noindent
\label{bayesFilter}
\hat{p}^{t+1}(s) = \frac{h(s,a^t,o^{t+1})\sum_{s' \in S }P(s',a^t,s)\hat{p}^t(s')}{\sum_{\sigma \in S} h(\sigma,a^t,o^{t+1})\sum_{s' \in S }P(s',a^t,\sigma)\hat{p}^t(s')}
\end{equation}
initialized with the prior distribution $\hat{p}^0$.

A \em deterministic transition system \em \cite{Baier2008}  is a tuple $TS =  (Q,q_0,Act, Trans, AP, L)$, where $Q$ is a set of states,
$q_0 \in Q$ is the initial state, $Act$ is a set of actions, $AP$ is a set of atomic propositions, $L:Q \to 2^{AP}$ is a mapping from states to propositions, and $Trans \subseteq Q \times Act \times Q$ is a transition relation such that $(q,a,q') \in Trans$ means performing action $a$ drives the state of $TS$ from $q$ to $q'$.  A finite word $a^{0:n} \in {Act}^{*}$ defines a run $q^{0:n} \in Q^{*}$ such that $q^0=q_0$ and $(q^i,a^i,q^{i+1}) \in Trans$ $\forall i = 1,2, \ldots, n-1$.

In this paper, we use syntactically co-safe linear TL (scLTL) as a basis for the definition of a new temporal logic.  An scLTL formula is inductively defined as follows \cite{Kupferman2001}:
\begin{equation}
	\phi := \pi | \neg \pi | \phi \vee \phi |  \phi \wedge \phi |  \phi \LTLUNTIL \phi | \LTLNEXT \phi | \LTLEVENTUALLY \phi,
\end{equation}
where $\pi$ is an atomic proposition, $\neg$ (negation), $\vee$ (disjunction), and $\wedge$ (conjunction) are Boolean operators, and $\LTLNEXT$ (``next''), $\LTLUNTIL$ (``until''), and $\LTLEVENTUALLY$ (``eventually'') are temporal operators.

A  \em (deterministic) finite state automaton \em (FSA) is a tuple  $\mathcal{A} = (\Sigma, \Pi, \Sigma_0, F, \Delta_\mathcal{A})$ where
$\Sigma$ is a finite set of states, $\Pi$ is an input alphabet, $\Sigma_0 \subseteq \Sigma$ is a set of initial states, $F \subseteq \Sigma$ is a set of final (accepting) states, and $\Delta_\mathcal{A} \subseteq \Sigma \times \Pi \times \Sigma$ is a deterministic transition relation.

An \em accepting run \em $\ARUN_\mathcal{A}$ of an automaton $\mathcal{A}$ on a finite word $w=w^0w^1 \ldots w^j$ over $\Pi$ is a sequence of states $\ARUN_\mathcal{A}=\sigma^0\sigma^1 \ldots \sigma^{j+1}$ such that  $\sigma^{j+1} \in F$ and $(\sigma^i,w^i,\sigma^{i+1})  \in \Delta_\mathcal{A}$ $\forall i \in [0,j]$. 
 
 Given an scLTL formula $\phi$, there exist algorithms for creating an FSA that accepts only words satisfying $\phi$ and there are known procedures for using such an FSA to check deterministic \cite{Latvala2003} or probabilistic \cite{Ulusoy2012c} models for satisfaction of $\phi$.  

\section{Motivating Example: Hypothesis Testing}
\label{related}
In this section, we use a simple multiple hypothesis testing example to motivate the introduction of the logic scLDTL described in Section \ref{scLDTL}.  Consider an experiment  in which one of three coins, each with different expected frequency of heads, is flipped repeatedly. This is an example of a Hidden Markov Model (HMM).  The hidden states of the system are $S_h = \{s_1,s_2,s_3\}$, where $s_i$ is a coin with heads frequency $p_i$.  The set of observations is $Obs = \{o_1,o_2\}$ where $o_1$ is heads and $o_2$ is tails.

\begin{figure}
 \begin{center}
  \includegraphics[scale=0.85]{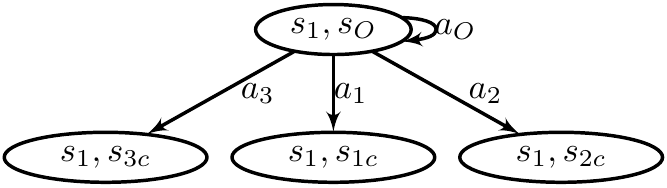}
  \caption{\label{Agent} A representation of the hidden state dynamics of the multiple hypothesis testing POMDP given that the true source is $s_1$.  The full state dynamics is given by three separate graphs of the same form where the first element $s_i$ in the state tuple indicates the true source of the observation sequence. }
 \end{center}
\end{figure}

Further, consider a deciding agent that at each time step can either make an observation from the HMM or choose a hypothesis in $S_h$.  Let $ S = S_h \times S_d$, where $S_d = \{s_{1c},s_{2c},s_{3c},s_O\}$ is the state space of the deciding agent. $s_O$ means that the HMM is being observed and $s_{ic}$ means that the hidden state $s_i$ is chosen as the most likely hypothesis.  The process is illustrated in Figure \ref{Agent}.
Combining the observation model from the HMM with the state dynamics described by Figure \ref{Agent} gives a POMDP $MHT=(S,\hat{p}^0,P,\{a_O,a_1,a_2,a_3\},\{o_1,o_2\},h)$ where $P$  and $h$ are given by
\begin{subequations}
\begin{equation}
 \begin{array}{l l}
  P([s_i,s_0],a_0,[s_i,s_0]) = 1, &  \\ P([s_i,s_0],a_j,[s_i,s_{jc}]) = 1 & \forall i,j \in \{1,2,3\} \\
  P(s,a,s') = 0,  & \text{ otherwise }
 \end{array}
\end{equation}
\begin{equation}
\begin{array}{l l}
 h([s_i,s_O],a_O,o_1) = p_i, & h([s_i,s_O],a_O,o_2) = 1-p_i  \\ h(s,a,o) = 0, & \text{ otherwise }
 \end{array}
\end{equation}
\end{subequations}
  
Consider the problem in which we are given an infinite number of observations from $MHT$, but must estimate the state of the system in finite time. One solution method is to prescribe a threshold on the entropy of the belief state and terminate observation and select the most likely hypothesis  when it is reached.  In plain English, this is ``When the entropy of the belief state is below $h$, select the most likely hypothesis." 

This can easily be described by the new Distribution Temporal Logic (DTL) we define in Section \ref{scLDTL}.  As it will become clear later, this predicate logic is defined over two types of predicates: belief predicates and state predicates.  ``When the entropy of the belief state is below $h$'' is equivalent to the belief predicate $H(\hat{p})-h<0$, which can be written in short as $(H(\hat{p})-h)$ where $H(\cdot)$ denotes entropy.  ``The most likely hypothesis'' is equivalent to $s_i$ such that $\hat{p}([s_i,s_O]) > \hat{p}([s,s_O])$ $ \forall s \in S_h \setminus \{s_i\}$.  Each comparison between components of $\hat{p}$ is a belief predicate.  The selection of hypothesis $s_i$ means the state is in the set $\{[s_j,s_{ic}]\}_{j \in \{1,2,3\}}$, and such sets will be referred to as state predicates.   As it will become clear in Section \ref{scLDTL}, the overall specification translates to the following DTL formula
\begin{equation}
 \label{exampleSpec}
 \begin{array}{l}
(H(\hat{p})-h) \Rightarrow  \\ (\bigwedge_{s_i \in S_{h}} (\bigwedge_{s_j \in S_{h} \setminus \{s_i\}} (\hat{p}([s_j,s_O]) - \hat{p}([s_i,s_O]))  \Rightarrow \\ \LTLNEXT \{[s_j,s_{ic}]\}_{j \in \{1,2,3\}}),
\end{array}
\end{equation}
where the temporal and logical operators have roughly the same semantics as scLTL ( see Section \ref{scLDTL}, Definition \ref{realizationBelief}).  

Neither the threshold on entropy nor the selection of the most likely hypothesis can be formulated using POCTL*, the existing temporal logic for partially observable Markov chains \cite{Zhang2004}.   POCTL* can describe some properties with respect to a belief state, namely whether the probability under the initial belief state of a collection of sample paths of hidden states and observations occurring is greater than or less than some threshold, but this calculation is a linear function of the belief state.   As entropy is a non-linear function of the belief state, entropy levels cannot be described in POCTL*.

The collection of sample paths that can be produced by the hidden state of the system are infinite repetitions of the $s_i$.  The probability of sample path $s_i s_i \ldots$ under a belief state is $\hat{p}([s_i,s_O])$.  In POCTL* for this problem, we can only compare the probability under a belief state of a single hypothesis or a pair of hypotheses to a constant value: we cannot compare the estimated probabilities of hypotheses to each other.  Therefore, we cannot use POCTL* to formulate the selection of the most likely hypothesis.

Since the problem we consider here is readily addressed with tools from optimal estimation and information theory (see e.g. \cite{Cover2006,Scharf1991}), constructing a new TL to describe the solution strategy may seem unnecessary. However, even considering only belief predicates that describe measures of uncertainty allows the description of novel behaviors.  We can specify low uncertainty levels as temporal goals.  We can use uncertainty thresholds to trigger behavior consistent with the most likely state(s) of the POMDP.  
Consider an agent tasked with target localization in a cluttered environment.  If the agent determines that an object is an obstacle, it must then avoid it.  If the agent determines the object is a target, it must return to base to report the location.

\section{Syntactically co-safe linear distribution temporal logic}
\label{scLDTL}

Syntactically co-safe linear distribution temporal logic (scLDTL) describes co-safe temporal logic properties of probabilistic systems and is defined over two types of predicates: belief predicates of the type $f<0$, with $f\in F_S:\{f:Dist(S) \to \mathbb{R}\}$ (denoted simply by $f$) where $Dist(S)$ is the set of all pmfs that can be defined over state space $S$ and state predicates $s\in A$, with $A\in 2^S$ (denoted simply by $A$). Formally, we have:
\begin{dfn}[scLDTL syntax]
\label{syntax}
An scLDTL formula over predicates over $F_{S}$ and state sets is inductively defined as follows:
\begin{equation}
	\phi := A | \neg A |  f | \neg f | \phi \vee \phi |  \phi \wedge \phi |  \phi \LTLUNTIL \phi | \LTLNEXT \phi | \LTLEVENTUALLY \phi,
\end{equation}
where $A \in 2^{S}$ is a set of states, $f \in F_{S}$ is a belief predicate, $\phi$ is an scLDTL formula, and $\neg$, $\vee$, $\wedge$, $\LTLNEXT$, $\LTLUNTIL$ , and $\LTLEVENTUALLY$ are as described in Section \ref{prelims}.
\end{dfn}
As scLDTL is defined over state and belief predicates, we construct a basic notion of satisfaction over pairs of hidden state sample paths and sequences of belief states, given by Definition \ref{realizationBelief}.
\begin{dfn}[scLDTL satisfaction of sample path/belief state sequence pairs]
\label{realizationBelief}
The semantics of scLDTL formulae is defined over words $w \in (S \times Dist(S))^{\infty}$.  Denote the $i$th letter in $w$ as $(s^i,\hat{p}^i)$
The satisfaction of a scLDTL formula at position $i$ in $w$, denoted  $(s^i,\hat{p}^i) \models \phi $, is recursively defined as follows:

\begin{itemize}
\item $(s^i,\hat{p}^i) \models A$ if $ s^i \in A$, 
\item $(s^i,\hat{p}^i) \models f$ if $ f(\hat{p}^i) <0$,
\item $(s^i,\hat{p}^i) \models \neg A$ if $ s^i \not \in A$, 
\item $(s^i,\hat{p}^i) \models \neg f$ if $ f(\hat{p}^i)  \geq 0$,
\item $(s^i,\hat{p}^i) \models \phi_1 \wedge \phi_2$ if $(s^i,\hat{p}^i) \models \phi_1$ and $(s^i,\hat{p}^i) \models \phi_2$, 
\item $(s^i,\hat{p}^i) \models \phi_1 \vee \phi_2$ if $(s^i,\hat{p}^i) \models \phi_1$ or $(s^i,\hat{p}^i) \models \phi_2$, 
\item $(s^i,\hat{p}^i) \models \LTLNEXT  \phi$ if $(s^{i+1},\hat{p}^{i+1}) \models \phi$, 
\item $(s^i,\hat{p}^i) \models \phi_1 \LTLUNTIL \phi_2$ if there exists $j \geq i$ such that $(s^j,\hat{p}^j) \models \phi_2$ and for all $i \leq k < j$ $(s^k,\hat{p}^k) \models \phi_1$,
\item $(s^i,\hat{p}^i) \models \LTLEVENTUALLY \phi$ if there exists $j \geq i$ such that $(s^j,\hat{p}^j) \models \phi$.
\end{itemize}

The word $w \models \phi$, iff $(s^0,\hat{p}^0) \models \phi$.
\end{dfn}

We also define a notion of probabilistic satisfaction with respect to an execution of a POMDP in Definition \ref{probSatisfy}.
\begin{dfn}[scLDTL satisfaction with respect to a POMDP execution]
\label{probSatisfy}
An execution of a POMDP (a sequence of belief states $\hat{p}^{0:t}$, the sequence of actions taken $a^{0:t-1}$, and the sequence of observations seen $o^{1:t}$)  \em probabilistically satisfies \em the scLDTL formula $\phi$ with probability $Pr [\{s^{0:t} \text{ such that }  (s^0,\hat{p}^0)\ldots(s^t,\hat{p}^t) \models \phi \}|\hat{p}^{0:t},a^{0:t-1},o^{1:t}]$, denoted in shorthand as $Pr[\phi|\hat{p}^{0:t},a^{0:t-1},o^{1:t}]$.
\end{dfn}

The probability of a single sample path conditioned on a POMDP execution may be calculated via the process of recursive smoothing \cite{Briers2010}. Note that we define $Pr[\phi|\hat{p}^{0:t},a^{0:t-1},o^{1:t}]$ with respect to finite-length sample paths.  Although the semantics of scLDTL is defined over infinite words, it is known that any co-safe temporal logic formula can be checked for satisfaction in finite time \cite{Latvala2003}.

\section{Monitoring POMDPs}
\label{accept}
Here we show algorithmically how to solve the following problem.

\begin{prob}[scLDTL monitoring of POMDPs]
 Evaluate with what probability a given finite-length execution of a POMDP $POMDP = (S,\hat{p}^0,P,Act,Obs,h)$ satisfies an scLDTL formula $\phi$ over subsets of $S$ and belief states over $S$.
\end{prob}
 
The solution to this problem could be used to evaluate the performance of a single execution of a POMDP or, as we show in Section \ref{caseStudy}, can be used to compare the performance of control policies.  More importantly, the tools developed for this problem are potentially useful for developing synthesis procedures.

The evaluation proceeds in two stages.  In the first stage, called feasibility checking, we check a necessary condition for the given execution to satisfy $\phi$ with $Pr[\phi|\hat{p}^{0:t},a^{0:t-1},o^{1:t}] > 0$. The second stage is probabilistic satisfaction checking, in which $Pr[\phi|\hat{p}^{0:t},a^{0:t-1},o^{1:t}]$ is calculated if feasibility checking has succeeded.

\subsection{Feasibility checking}

 Algorithm \ref{sysConst} shows how to construct a deterministic transition system whose labels correspond to the belief predicates involved in the scLDTL formula $\phi$.  In order to incorporate the state predicates into this discrete system, we  relax all state predicates by mapping them to belief predicates, e.g., state predicate $A$ is relaxed to the belief predicate $(-\sum_{s \in A} \hat{p}(s))$ (i.e. $Pr[s \in A] > 0$) (line \ref{relax}).  We also create a mapping $\Psi_{F}$ from each belief predicate to an atomic proposition (lines \ref{gatherStart}-\ref{gatherEnd}).  Then, for each $f$ appearing in the relaxed scLDTL formula, we calculate the level set $f(\hat{p})=0$ in $Dist(S)$ and map it to a set of probability vectors in the probability simplex. Many useful belief predicates, such as inequalities over moments, have polytopic level sets that are readily calculated. The level sets induce a partition of the simplex. A general algorithm for producing this partition will likely require the use of 
geometric tools and direct evaluations of the functions $f$ for points in the simplex.  We take the quotient of the partition to form a transition system and label each state with $\Psi_{F}(f)$ for each $f$ that was satisfied in the 
corresponding region (lines \ref{stateStart}-\ref{systemEnd}).  We denote the region of the simplex corresponding to the state $q_j$ in the transition system as $Reg(q_j)$.

The condition in line \ref{reach} used to create transitions in the quotient involves a notion of reachability that we make precise now.

\begin{dfn}[Reachability]
\label{finiteReach}
We say a state $q_k$ is reachable from state $q_m$ if beginning from any belief state in $Reg(q_m)$ there exists a sequence of actions and observations in $POMDP$ such that sequentially applying \eqref{bayesFilter} will drive the system to a belief state associated with a belief state in $Reg(q_k)$.
\end{dfn}

Determining the reachability relationship between states is a non-trivial process. In this work, we assume that all states are self-reachable and all state pairs corresponding to neighboring regions in the probability simplex are mutually reachable. We make this liberal assumption because if we observe a transition during monitoring that we did not assume to exist, $FTS$ would be invalid.  Allowing all possible transitions does not weaken our approach if a reachability relationship is false. If a transition cannot be made in $FTS$, we will never observe it during monitoring. This assumption will have to be relaxed in model checking or synthesis. Further, each transition is annotated with a virtual action rather than a collection of action/observation sequences.

\begin{algorithm}
\caption{Construct a transition system used to check a necessary condition for $Pr[\phi|\hat{p}^{0:t},a^{0:t-1},o^{1:t}]>0$}
\label{sysConst}
\begin{algorithmic}[1]
\FUNCTION{feasibilitySystemConstruct}{$\phi,S$,$\hat{p}^0$} \label{beg}
\STATE predicateSet := $\emptyset$ ; $j$ := 1; $\Pi = \emptyset$;
\FORALL {predicates $\in \phi$} \label{gatherStart}
	\IF {predicate $\not \in F_{S}$}
		\STATE predicate := $(-\sum_{s \in \text{predicate}} \hat{p}(s))$  \label{relax}
		\ENDIF
	\STATE predicateSet := predicateSet $\cup$ predicate
	\STATE $\Psi_{F}$(predicate) := $\pi_j$ \label{gatherEnd} 
	\STATE $\Pi := \Pi \cup \pi_j$; $j := j+ 1$;\ENDFOR
\FORALL{$f \in $ predicates} 
\STATE calculate level set$f(\hat{p}) = 0$ \ENDFOR
\STATE use the probability vector representation of the level sets to partition the probability simplex
\STATE $Q_{F} := \emptyset$; m := 1;
\FORALL{regions $\in$ partition} \label{stateStart}
	\STATE $Q_{F} := Q_{F} \cup \{q_m\}$;
	\STATE $L_{F}(q_{m}) := \{\Psi(f)|f(\hat{p}) < 0 $ $\forall \hat{p} \in $ region$\}$
	\STATE $Reg(q_m)$ := region
	\IF{$\hat{p}^0 \in$ region}
		\STATE $q_0 := q_m$; \ENDIF
	\STATE $ m := m+1$ \label{stateEnd}
\ENDFOR 
\STATE $Act_{F} := \emptyset;$ $Trans_{F} := \emptyset$
\FORALL{$q_m,q_k \in Q_{F}^2$} \label{systemStart}
	\IF{$q_k$ is reachable from $q_m$} \label{reach}
		\STATE $Act_{F} := Act_{F} \cup \{a_{mk} \}$
		\STATE $Trans_{F} := Trans_{F} \cup \{(q_m, a_{mk}, q_k )\}$ \label{systemEnd}
	\ENDIF \ENDFOR 
\RETURN{$FTS = (Q_{F},q_{0},Act_{F},Trans_{F},\Pi_{F},L_{F}),\Psi_{F}$}
\end{algorithmic}
\end{algorithm}

Feasibility checking of a scLDTL formula proceeds according to Algorithm \ref{feasCheck}. 
From $\phi$, we create an scLTL formula $\phi'$ by replacing every predicate in $\phi$ with its image in the mapping $\Psi_{F}$ (lines \ref{mapStart} - \ref{mapEnd}).  We then construct the automaton $\mathcal{A}_{\phi'}$ and form $\mathcal{P}_{\phi'}$, the synchronous product  of $FTS$  (from Algorithm \ref{sysConst}) and $\mathcal{A}_{\phi'}$.  The sequence $\hat{p}^{0:t}$ is translated into the corresponding word $\alpha^{0:t}$ in the input language of $\mathcal{P}_{\phi'}$ (lines \ref{inputStart} - \ref{inputEnd}).  We use $\mathcal{P}_{\phi'}$ to perform scLTL verification of $\phi'$.  If verification succeeds, Algorithm \ref{feasCheck} returns a deterministic transition system $DTS$  used in probabilistic acceptance checking to describe the time evolution of the satisfaction of belief predicates.  
$DTS$ is a simple,``linear'' transition system whose action set is a singleton and whose only possible run  is $q_0 \ldots q_t$ where $L_{D}(q_k) = L_{F}(q | \hat{p}^k \in Reg(q))$.


\begin{algorithm}
\caption{Returns a transition system that describes the time evolution of belief predicate satisfaction if the necessary condition for $Pr[\phi|\hat{p}^{0:t},a^{0:t-1},o^{1:t}]>0$ holds}
\label{feasCheck}
\begin{algorithmic}[1]
\FUNCTION{feasibilityCheck}{$\hat{p}^{0:t}, \phi,S$} 
\STATE $FTS,\Psi_{F}$ := feasibilitySystemConstruct($\phi,S$,$b^0$) 
\STATE $\phi' : = \phi$ \label{mapStart}
\FORALL {predicates $\in \phi'$}
	\STATE replace predicate in $\phi'$ with $\Psi_{F}$(predicate); \label{mapEnd}
	\ENDFOR
\STATE Construct the finite state automaton (FSA) $\mathcal{A}_{\phi'}$ that only accepts words satisfying $\phi'$.
\STATE $\mathcal{P}_{\phi'} = FTS \times A_{\phi'}$
\STATE  currentState := $q_0$; currentIndex := 0; $k := 1$
\STATE $Q_{D} := \emptyset$; $Act_{D} = \{a_0\}$; $Trans_{D} = \emptyset$; $\Pi_{D} := L_{F}(q_0)$ \label{inputStart}
\FOR{ $i = 1$ to $t$}
	\IF{$ \hat{p}^i \not \in Reg($currentState)}
		\STATE currentState := $q_j$ such that $\hat{p}^i \in Reg(q_j)$
		\STATE $\Pi_{D} := \Pi_{D} \cup L_{F}($currentState$)$
		 \STATE currentIndex := $j$; \label{inputEnd}
		\ENDIF
	\STATE $Q_{D} := Q_{D} \cup q_k$
	\STATE $Trans_{D} := Trans \cup (q_{k-1},a_0,q_k)$
	\STATE $L_{D}(q_k) = L_{F}($currentState$)$
	\STATE $\alpha^i := a_{\text{currentIndex},\text{nextIndex}}$
	\STATE currentIndex := nextIndex
	\ENDFOR
\IF{ $\alpha^{0:t-1}$ produces an accepting run on $\mathcal{P}_{\phi'}$} \label{prodCond}
	\RETURN{$DTS = (Q_{D},q_{0,D},Act_{D},Trans_{D},\Pi_{D},L_{D}$} 
	\ENDIF
\RETURN{False} \label{verEnd}
\end{algorithmic}
\end{algorithm}

 If verification fails, then we do not proceed to probabilistic acceptance checking, as failure means that $Pr[\phi|\hat{p}^{0:t},a^{0:t-1},o^{1:t}] = 0$.  Due to the mapping of state predicates to belief predicates, Algorithm \ref{feasCheck} checks for the existence of at least one sample path $s^{0:t}$ such that $(s^0,\hat{p}^0) \ldots (s^t,\hat{p}^t) \models \phi$ and $\prod_{i=0}^t\hat{p}^i(s^i) > 0$.  The positivity of the product is a necessary but not sufficient condition for $Pr[s^{0:t}|a^{0:t-1},o^{1:t},\hat{p}^0] > 0$.
 
We illustrate Algorithms \ref{sysConst} and \ref{feasCheck} in the following example.
\begin{example}
Consider the multiple hypothesis testing POMDP $MHT$ given in Section \ref{related} with scLDTL specification \eqref{exampleSpec}. Figure \ref{ProbSimplex}(a) shows the partitioning of the probability simplex from the belief predicates in \eqref{exampleSpec}  resulting from Algorithm \ref{sysConst}.  The predicates involving maximum likelihood (red) and specified entropy level  (blue) partition the simplex into six regions corresponding to discrete states $q_i, i \in \{1,\ldots,6\}$.  Each red curve is a level set $\hat{p}([s_i,s_O]) = \hat{p}([s_j,s_0]$ for $i \neq j$ and each blue curve is part of the level set $H(\hat{p})=0.8$ bits.  From this partition, we can execute Algorithm \ref{sysConst}, lines \ref{stateStart}-\ref{systemEnd} to form the transition system $FTS$ shown in Figure \ref{ProbSimplex}(b).  A state in $FTS$ is labeled with proposition $\pi_j, j \in \{1,2,3\}$ if $s_j$ is the most likely hypothesis according to any probability vector in the corresponding region.  A state in $FTS$ is labeled 
with proposition $\pi_4$ if the entropy of any probability vector in 
that region is less than $0.8$ bits.

The green curve in Figure \ref{ProbSimplex}(a) represents a single, randomly generated execution of $MHT$.  The observation likelihood parameters were $p_1=0.25, p_2 =0.5, p_3=0.75$.  Observations were generated with parameter $p_1$. Each point in the curve is the probability vector representation of the belief state $\hat{p}^i$ resulting from incorporating $i$ observations via \eqref{bayesFilter}. The transition system $DTS$ resulting from executing Algorithm \ref{feasCheck} on the given sequence of belief states is shown in Figure \ref{ProbSimplex}(c).  For the first three observations seen, the trajectory stays in $Reg(q_1)$.  Thus the first three states in $DTS$ are labeled with $\pi_1$.  After the fourth measurement, the trajectory  has gathered enough information to enter $Reg(q_4)$.  Thus the fourth (and final) state in $DTS$ is labeled with both $\pi_1$ and $\pi_4$.   

\begin{figure*}
 \begin{center}
 \begin{tabular}{c c c}
  \includegraphics[width=0.3\textwidth]{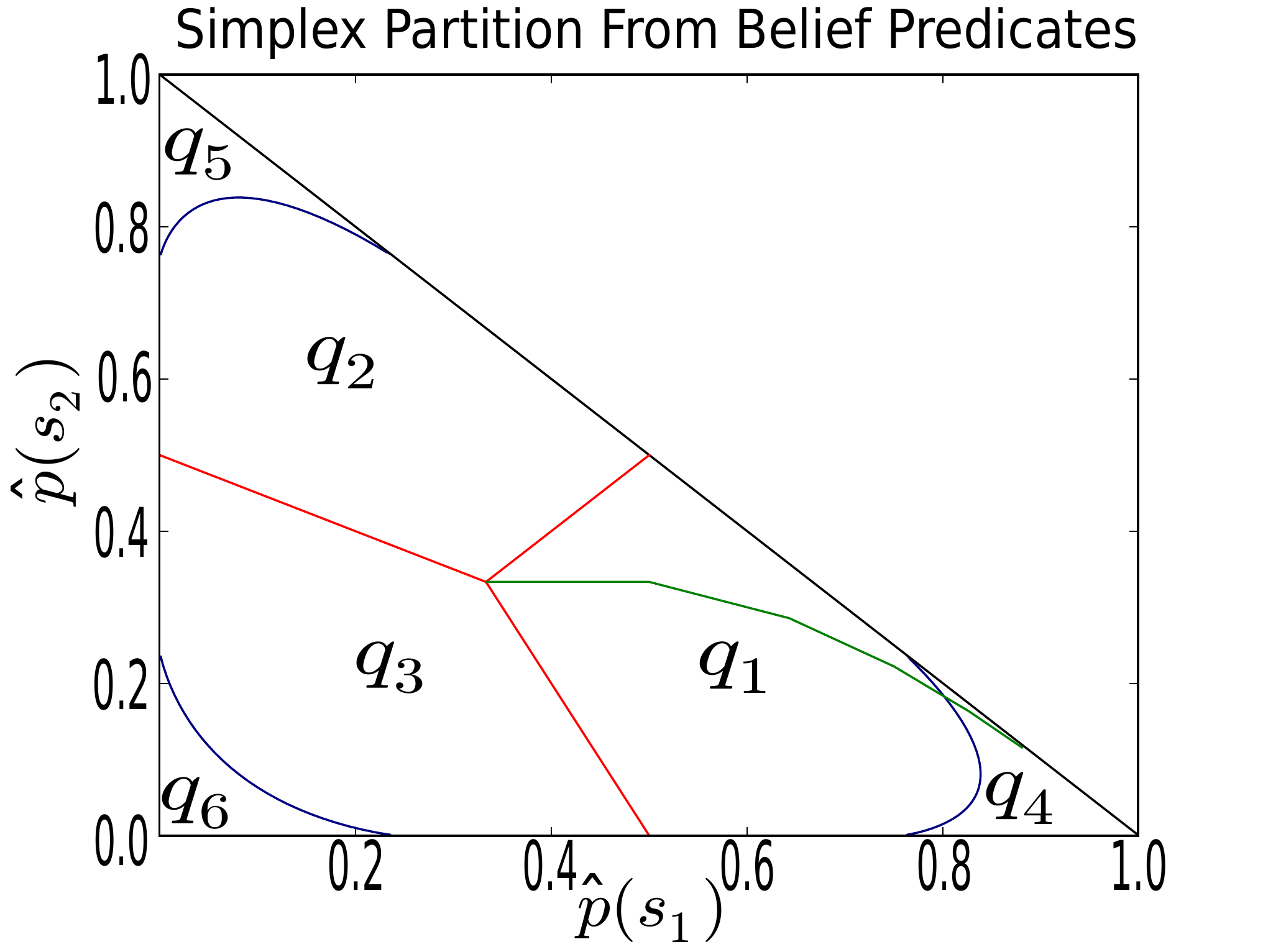} & 
  \includegraphics[width=0.35\textwidth]{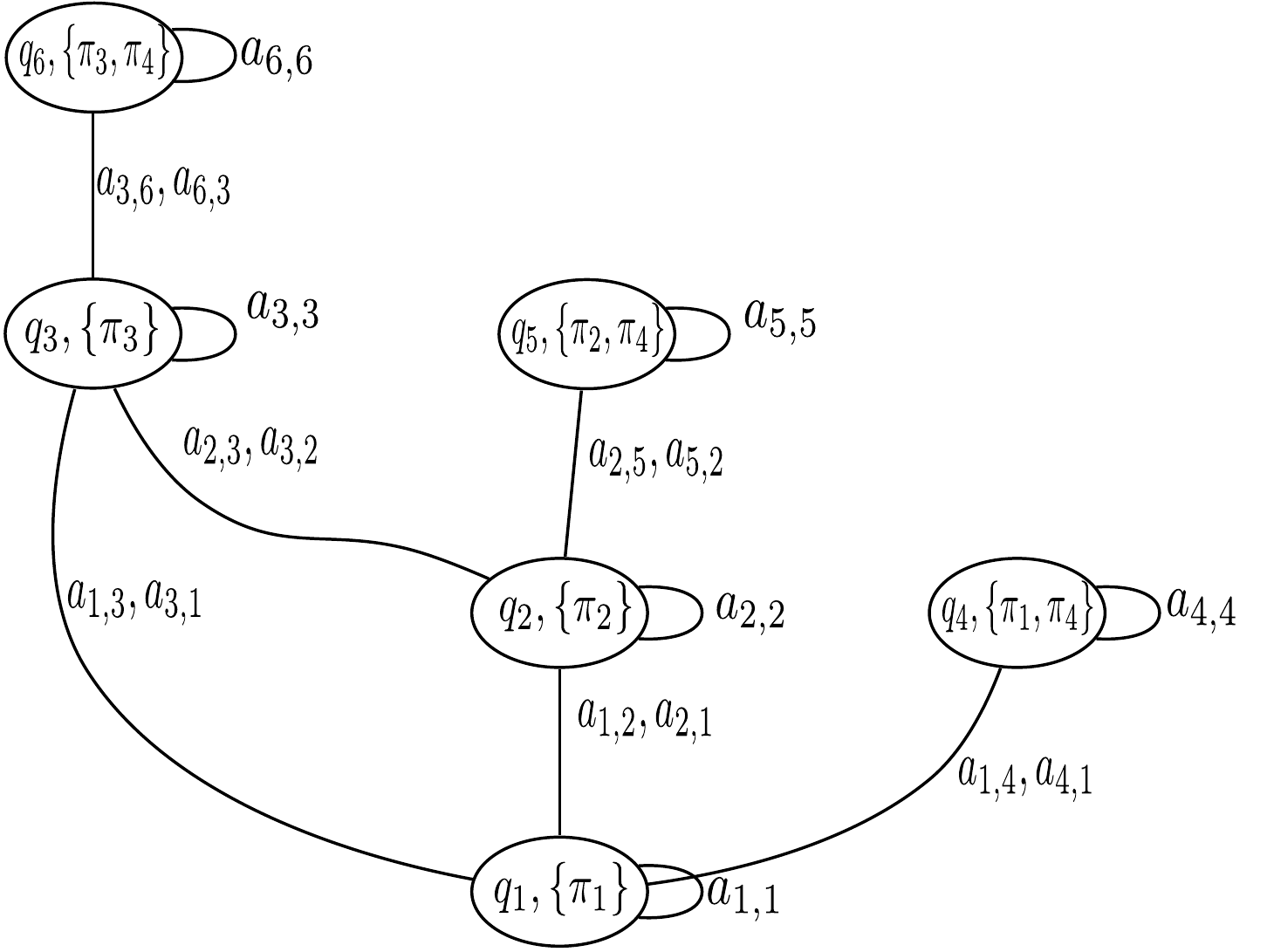} &
  \includegraphics[width=0.25\textwidth]{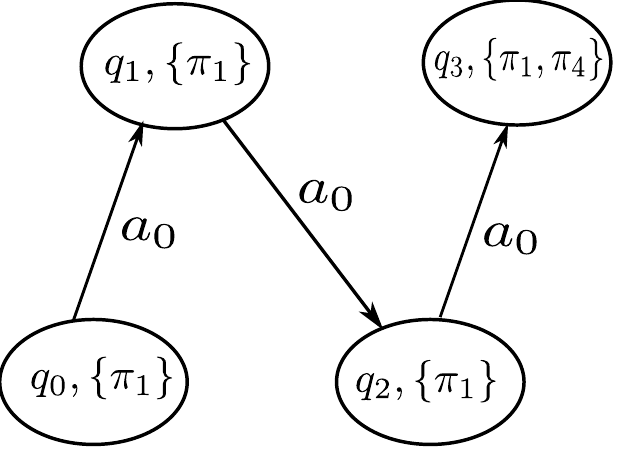} \\
  (a) &(b) & (c) \\
  \end{tabular}
 \caption{\label{ProbSimplex} (a) The probability simplex for $\hat{p}_{S_h}$ partitioned according to the belief predicates used in \eqref{exampleSpec}. The red lines divide the simplex into three regions corresponding to the most likely hypothesis.  The blue curves are the level sets $H(\hat{p}_S) = 0.8$ bits.  The green curve shows the probability trajectory corresponding to a sequence of belief states from a randomly generated execution of $MHT$. (b) The transition system $FTS$ constructed by taking the quotient of the partition shown in (a).  (c)  The transition system $DTS$ that results from applying Algorithm \ref{feasCheck} to the given belief state sequence and $FTS$.}
\end{center}
\end{figure*}

\end{example}

\subsection{Probabilistic acceptance checking}
\label{probAccept}

If Algorithm \ref{feasCheck} succeeds, we proceed to probabilistic acceptance checking.  In this section, we use labeled Markov decision processes (LMDPs) and labeled Markov chains (LMCs) as abstractions to describe the probabilistic time evolution of the hidden states of the system. An LMDP is a POMDP in which the states of the system are fully observable and labeled with atomic propositions. An LMDP is given as a tuple $LMDP = (S,p_S^0,P,Act,AP,L)$ where $S,P,$ and $Act$ are as defined for a POMDP.  The pmf over states $p_S$ is not conditioned on observations.  $AP$ is a set of atomic propositions and $L:S \to 2^{AP}$ maps states to propositions. A labeled Markov Chain (LMC) is an LMDP without actions and is given as a tuple $LMC = (S,p_S^0,P,AP,L)$ where $S,p_S^0,AP,$ and $L$ are as defined for the LMDP and the probabilistic transition relationship $P$ is not parameterized by actions.

We begin probabilistic acceptance checking by  creating a mapping $\Psi_{sp}: 2^S \to \Pi_r$ that maps state predicates to atomic propositions in the set $\Pi_r$.  This construction is similar to the construction of $\Psi_F$.  The scLDTL formula $\phi$ is mapped to a scLTL formula $\phi''$ by applying the mapping $\Psi_{F}$ to the belief predicates and the mapping $\Psi_{sp}$ to the state predicates appearing in $\phi$.  An FSA is created from $\phi''$. Next, we enumerate all of the sample paths consistent with the given execution of $POMDP$.  We do this by creating a labeled Markov chain $LMC$ for each possible initial state $s^0$ such that $\hat{p}^0(s^0) > 0$.  $LMC$ has a tree-like structure with root $s^0$.  Each node  $s^i$ has as children any state $s^{i+1}$ such that $P(s^{i},a^{i},s^{i+1}) > 0$ and $h(s^{i+1},a^i,o^{i+1}) >0$.  Each state $s$ in the tree is labeled with $\{\Psi_{sp}(A)| A$ appears in $\phi$, $s \in A\}$.  The transition probability between states $s^i,s^{i+1}$ is given by 
\begin{equation} \nonumber \noindent
 P_{LMC}(s^i,s^{i+1}) = \frac{Pr[o^{i+1:t}|s^{i+1},a^{i+1:t}]P(s^i,a^i,s^{i+1})}{\sum_{s \in S} Pr[o^{i+1:t}|s,a^{i+1:t}]P(s^i,a^i,s)}.
\end{equation}
The details of this calculation can be found in \cite{Briers2010}.  We construct $LMDP$, the synchronous product of $LMC$ and  $DTS$, which  encapsulates the time evolution of both the satisfaction of state predicates (from $LMC$) and belief predicates (from $DTS$).  A state in the $i$th level of $LMDP$ is labeled with atomic propositions associated with the belief predicates satisfied by $\hat{p}^i$ and state predicates satisfied by a state $s^i$ that is reachable from state $s^0$ given the first $i$ actions and observations.

Since the action set $Act_{DTS}$ is a singleton, there is no notion of choice in the evolution of $DTS$ and thus no choice in the evolution of $LMDP$.  We create another labeled Markov chain $LMC_P$ from $LMDP$ by removing the action set and using the probabilistic transition relationship $P_{LMC_P}(s,s') = P_{LMDP}(s,a_0,s')$.  We then form $\mathcal{M}$, the synchronous product of  $LMC_P$ and $\mathcal{A}_{\phi''}$.  We perform model checking on $LMC_P$ to find the set of all accepting runs $Acc(\phi'')$ on $\mathcal{M}$ of length $t+1$.  Each run in $Acc(\phi'')$ corresponds to a sample path that satisfies $\phi$ when paired with $\hat{p}^{0:t}$.  For each run $r^{0 :t}$ in $Acc(\phi'')$, let $s^{0:t}$ be the corresponding sample path over $LMC_P$. We calculate $Pr[s^{0:t}|a^{0:t-1},o^{1:t}] 
= Pr[s^0|a^{0:t-1},o^{0:t}]\prod_{i=1}^{t}P_{LMC_P}(s^{i},s^{i+1})$ \cite{Briers2010}  and add it to the acceptance probability $Pr[\phi|\hat{p}^{0:t},a^{0:t-1},o^{1:t}]$.  By enumerating over all possible sample paths, we calculate the exact value of $Pr[\phi|\hat{p}^{0:t},a^{0:t-1},o^{1:t}]$. 

\section{Case Study: rescue robots}
\label{caseStudy}
A proposed use of mobile robots is to perform rescue operations in areas that are too hazardous for human rescuers.  A robot is deployed to a location such as an office building or school after a natural disaster  and is tasked with finding all human survivors in the environment and with moving any immobilized survivors to safe areas.  
The robot must learn survivor locations and safety profile of the building on-line by processing noisy measurements from its sensors.  The combination of on-line estimation and time-sensitive decision-making indicates that scLDTL is a good framework for describing the mission specification at a high level.

\subsection{Model}
For simplicity, we consider a rescue robot acting in a two room environment.  
We model the robot as a POMDP $Rescue = (S,\hat{p}^0,P,Act,Obs,h)$. The state of the system is given by a vector  $[s_q,s_O,s_{1,e},s_{2,e},s_{1,s},s_{2,s}]$ in the state space $S = \{1,2\} \times \{0,1\}^5$.  The element $s_q$ corresponds to the room in which the robot currently resides and  $s_O \in \{0,1\}$ corresponds to whether ($s_O=1$) or not( the robot is carrying an object($S_O=0)$. The elements $s_{i,e} \in \{0,1\}$ correspond to safety, i.e.  if $s_{i,e} = 1$, then room $i$ is safe to be occupied by a human.  The elements $s_{i,s} \in \{0,1\}$ correspond to survivor presence, i.e. if $s_{i,s} = 1$, a survivor is in room $i$. 

The robot can stay in its current room and measure its surroundings, switch to the other room, pick up an object, or put down an object.   Here we assume the motion model of the robot is deterministic, the safety of the environment is static, and the survivor locations change only if the robot moves a survivor. If the robot attempts to move a survivor, it fails with some probability $p_{fail}$.

If the robot takes action $Stay$, its sensors return observations in the set $Obs = \{0,1\}^2$.  The elements of $Obs$ are binary reports of the safety and survivor occupancy of the current room.  The sensor is parameterized by two independent false alarm and correct detection rates.

\subsection{Problem statement}
For convenience we establish the shorthand $\hat{p}_{j}(\sigma) = \sum_{\{s \in S|s_j = \sigma\}} \hat{p}(s)$ where $s_j$ is a component of an element of $S$.
 We wish to find and move all of the survivors in the given area to safe regions.  In order for the robot to be reasonably sure that this condition is met, it must be fairly certain about the state of the environment.  Therefore, we want the entropy of our estimate to be low, i.e.
 \begin{equation}
 \label{entReq}
 \forall i \in \{1,2\} \text{ } H(\hat{p}_{i,e}) < h_1\text{, } H(\hat{p}_{i,s}) < h_2.
 \end{equation}

Survivor safety is time-critical. We thus require ``If the robot is confident a survivor is in an unsafe location, move it to a safe location".  We encode confidence by saying "with a certain probability".

The statement that describes the rescue robotics application is ``Explore the environment and if the robot is in a state where it is sure with probability $p_1$ there is a survivor and with probability $p_2$ the state is unsafe, pick up the survivor, move to the other room and deposit the survivor.  Perform these actions until \eqref{entReq} and any identified survivors are in safe regions".   The above statement is encoded in the scLDTL formula  $\phi_1 \LTLUNTIL \phi_2$ where
\begin{equation}
\label{rescueDTL}
\begin{array}{c c}
\phi_1 = & \begin{array}{c} (\{s|s_{q} =j\} \wedge (p_1 - \hat{p}_j(s))  \wedge (p_2 - \hat{p}_{j,e}(0))) \\ \Rightarrow (\LTLNEXT (\{s|s_O =1\}  \LTLUNTIL  \{s|s_{q}  \neq j\}) \wedge \LTLNEXT \{s|s_{O} = 0 \} \end{array} \\ & \\
\phi_2 =&  \begin{array}{c} \bigwedge_{i \in \{1,2\}} (H(\hat{p}_{i,e}) -h_1)   \wedge (H(\hat{p}_{i,s}) - h_2) \\ \wedge (\{s|s_{i,e}=1\}) \wedge\{s|s_{i,s}=1\})  \vee  \{s|s_{i,s}=0\}) \\ \end{array}
\end{array}
\end{equation}

The formula $\phi_1$ encodes ``if the robot is in a state where it is sure with probability $p_1$ there is a survivor ($\{s|s_{q} =j\} \wedge (p_1 - \hat{p}_j(s))$) and with probability $p_2$ the state is unsafe ($p_2 - \hat{p}_{j,e}(0)$), pick up the survivor ($\{s|s_O =1\}$),   move to the other room  $ \{s|s_{q}  \neq j\})$ and deposit the survivor ($\{s|s_{O} = 0 \} $)."  The formula $\phi_2$ encodes  ``Perform these actions until \eqref{entReq} ($ \bigwedge_{i \in \{1,2\}} (H(\hat{p}_{i,e}) -h_1)   \wedge (H(\hat{p}_{i,s}) - h_2)$)  and any survivors are in safe regions ($(\{s|s_{i,e}=1\}) \wedge\{s|s_{i,s}=1\})  \vee  \{s|s_{i,s}=0\})$)."

Due to the time sensitive nature of survival, we consider the following time-constrained optimization problem.

\begin{equation}
\label{rescue}
\begin{array}{c}
\underset{a^{0:t}} \max E_{\{o^{1:t}\}}[ Pr[\phi_1 \LTLUNTIL \phi_2|\hat{p}^{0:t},a^{0:t-1},o^{1:t}]] \\
\end{array}
\end{equation}

\subsection{Acceptance checking}

We consider two separate strategies: time share and entropy cutoff.  In the time share strategy with parameter $a$, the robot switches rooms every $\lceil \frac{t}{a} \rceil$ observations.  In the entropy cutoff strategy with parameters $h_3,h_4,\rho$, the robot switches rooms when the entropy of the estimate of the safety and survivor presence of the current room dips below $h_3$ and $h_4$, respectively.  If the estimates of both rooms are of the specified certainty, the agent must wait $\rho$ time units before switching.  Both strategies include the reactive behavior of attempting to pick up survivors when they are found.

\begin{figure*}
\begin{center}
 \begin{tabular}{c c}
 \includegraphics[width=0.3\textwidth]{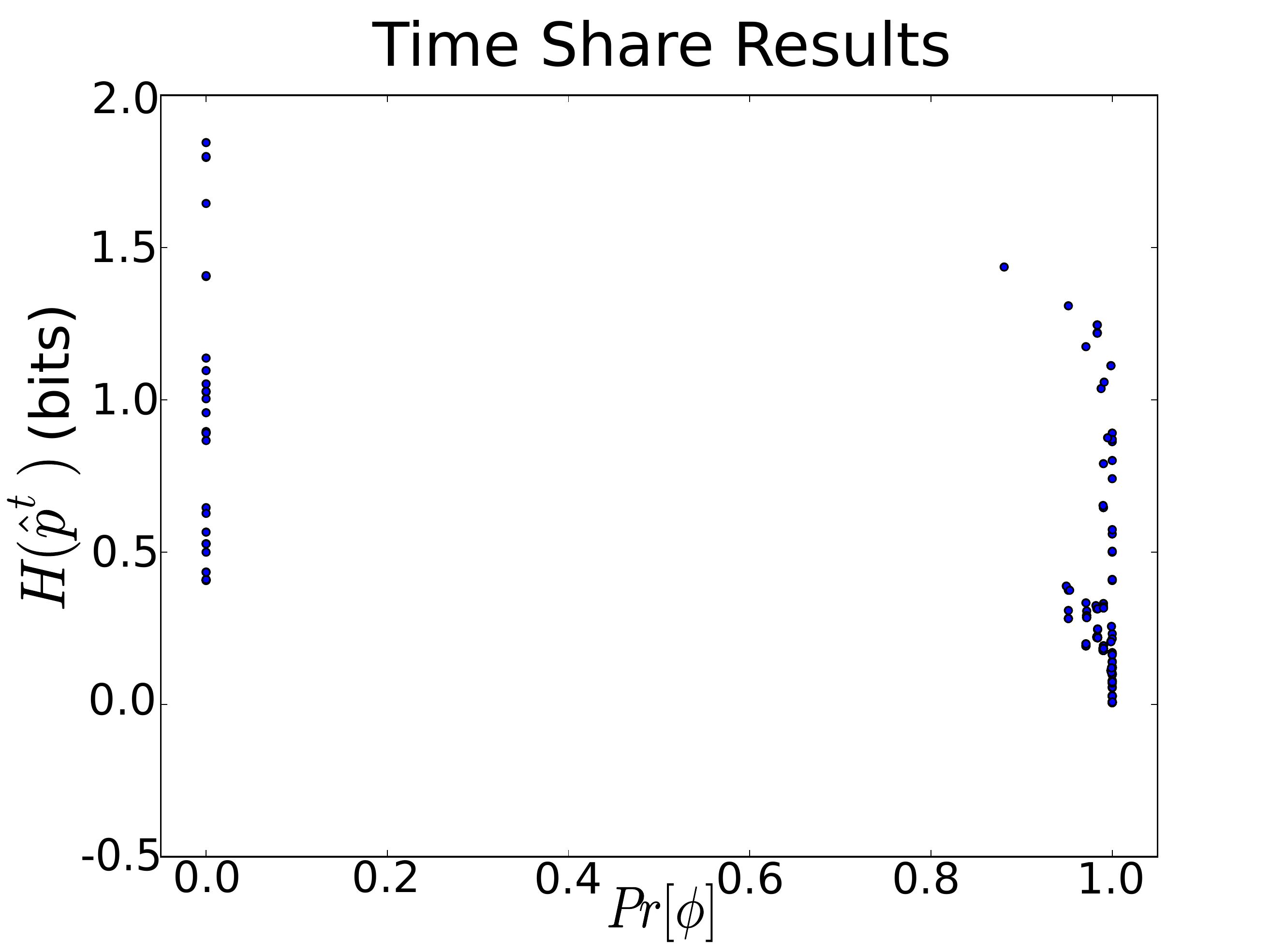}  & \includegraphics[width=0.3\textwidth]{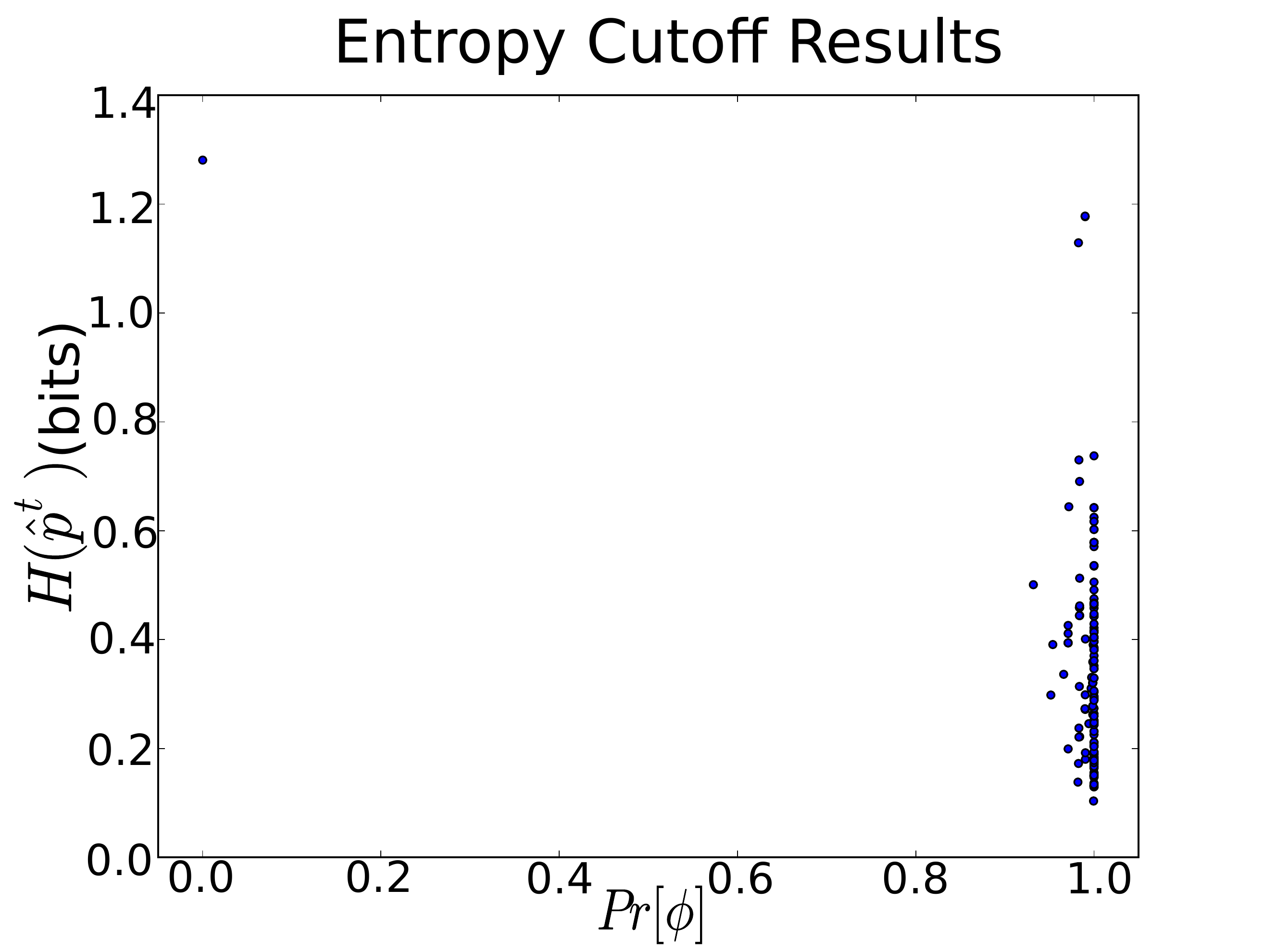}  \\ (a) & (b) \\
 \end{tabular}
 \caption{\label{scatters} Scatter plots showing the results of 250 Monte Carlo trials of the two room rescue robot POMDP under policy (a) time share and (b) entropy cutoff.  The parameters used in \eqref{rescueDTL} are $h_1=h_2=0.375$ and $p_1 = 0.9$, $p_2 = 0.25$.  The probability that an agent fails to pick up a survivor was $p_{fail} = 0.4$.  The false alarm rates for safety and survivor were both 0.1.  The correct detection rates for safety and survivor were 0.8 and 0.9, respectively.}
\end{center}
\end{figure*}

The results from 250 Monte Carlo trials of length $t=16$ are shown in Figure \ref{scatters}.  The control strategy parameters were parameter $a$=3,$h_3 = h_4 = 0.3$, and $\rho=2$. Further simulation parameters are given in the caption of Figure \ref{scatters}.  Here we use $Pr[\phi]$ as shorthand for the statistic formed from samples of $Pr[\phi|\hat{p}^{0:t},a^{0:t-1},o^{1:t}]$ collected from the trials.  For both methods, there are clusters of points around the lines $Pr[\phi]=1$ and $Pr[\phi]=0$.  This is because by making the entropy of the belief state a temporal goal in the scLDTL formula, the probability calculation sets the acceptance probability to 0 for executions after which the characterization of the environment is ambiguous, i.e. when the probability is close to the center of the interval $[0,1]$.  

The statistics resulting from our simulations are shown in Table \ref{stats}.  The statistic $r(Pr[\phi],H(\hat{p}^t))$ is the Pearson's $r$ correlation coefficient between the two variables.  The success rate is given as the number of trials such that at time $t=16$, all survivors were safe divided by the total number of trials. Note that the entropy cutoff method performs better in terms of acceptance probability, expected terminal entropy, and success rate. This matches intuition, as this method will  drive the robot to stay in a room longer if the particular observation sequence it observes does not lead to any strong conclusions or it will move to the other room if it has already obtained a good estimate.  This is in contrast to the time share method, which ignores estimate quality in its decision policy.

Further, note that for both methods, the correlation coefficient is weakly negative.  This weakness is due to the clustering of points at varying entropies around $Pr[\phi]=0$ and $Pr[\phi]=1$.  This negative correlation and the relative closeness of the average acceptance probability of the two methods to their respective success rates suggests that for some appropriately-defined scLDTL formulae, the probability $Pr[\phi|\hat{p}^{0:t},a^{0:t-1},o^{1:t}]$ is an appropriate metric for the dual consideration of estimate quality (uncertainty) and system performance.


\begin{table*}
\begin{center}
\begin{tabular}{|l|l|l|l|l|l|l|} 
\hline
Method & $E[Pr[\phi]]$ & $var(Pr[\phi])$ & $E[H(\hat{p}^t)]$ & $var(H(\hat{p}^t)$ & success rate & $r(Pr[\phi],H(\hat{p}^t))$ \\ 
\hline
Timeshare & 0.855 & 0.115 &  0.366 bits & 0.150 bits$^2$ &  0.86 & -0.547\\
\hline
Entropy Threshold & 0.992  & 0.004 & 0.338 bits & 0.034 bits$^2$ & 0.916 & -0.341 \\
\hline
\end{tabular}
\caption{\label{stats} Statistics from 250 Monte Carlo trials of the two-room rescue robotics simulation.}
\end{center}
\end{table*}

\section{Conclusions}
We argued that a new type of temporal logic, generically denoted as Distribution Temporal Logic (DTL), is needed to express notions of uncertainty and ambiguity in partially observed systems.  We have formalized a co-safe version of this logic and shown how to evaluate with what probability an execution of a POMDP satisfies a DTL formula.  Our case study demonstrates that this probability is a relevant metric for the performance of control policies.  In the future, we will extend these results to a procedure for synthesizing control policies that maximize this probability.  The application of DTL to other probabilistic systems and further exploration of its expressivity are also planned areas of research.

\bibliographystyle{plain}
\bibliography{../../../Literature/InfoTheoryFormalMethodsRobotics}

\end{document}